\documentclass[aps,superscriptaddress,eqsecnum,nofootinbib,preprintnumbers]{revtex4}

\usepackage{graphicx,epsfig}
\usepackage{amssymb,amsmath,bigints}
\usepackage{subfigure}

\usepackage{relsize}

\usepackage[utf8]{inputenc}

\usepackage{amsmath}
\usepackage{amssymb}

\DeclareMathOperator{\arcsinh}{arcsinh}

\newcommand*\di{\mathop{}\!\mathrm{d}}

\usepackage{hyperref}
\usepackage{url}
\usepackage{xcolor}
\usepackage{color}
 
\newcommand{\Regk}{\ensuremath{R_k}}
\definecolor{amaranth}{rgb}{0.9, 0.17, 0.31}
\definecolor{purple(munsell)}{rgb}{0.62, 0.0, 0.77}
\definecolor{americanrose}{rgb}{1.0, 0.01, 0.24}
\definecolor{palatinateblue}{rgb}{0.15, 0.23, 0.89}
\definecolor{royalblue(web)}{rgb}{0.25, 0.41, 0.88}
\definecolor{hanpurple}{rgb}{0.32, 0.09, 0.98}
\definecolor{beaublue}{rgb}{0.74, 0.83, 0.9}
\definecolor{carminered}{rgb}{1.0, 0.0, 0.22}
\definecolor{brightpink}{rgb}{1.0, 0.0, 0.5}
\definecolor{vividviolet}{rgb}{0.62, 0.0, 1.0}
\hypersetup{ linktoc=all,
    colorlinks, linkcolor={palatinateblue},
    citecolor={brightpink}, urlcolor={amaranth}}

\begin{document}

\title{Phenomenological footprints of $\Lambda$ varying gravity theories inspired from  quantum gravity models in the multi-messenger era}

\author{Michael R.R. Good}
\address{
Department of Physics \& Energetic Cosmos Laboratory, Nazarbaev University, Astana 010000, Qazaqstan}
\address{Leung Center for Cosmology and Particle Astrophysics, National Taiwan University,
Taipei 10617, Taiwan}

\author{Vasilios Zarikas}
\address{Department of Mathematics, University of Thessaly, 35100, Lamia, Greece}
\email{vzarikas@uth.gr}

\vspace{10pt}

\begin{abstract}

An interesting phenomenological consequence of $\Lambda$ varying gravity theories inspired by quantum gravity models is reported. 
The treatment in the present work is quite general and applicable to several different actions with $\Lambda$ varying, especially those used in RG approaches to quantum gravity. 
An effective gravitational action with a scale varying cosmological constant, $\Lambda$, which depends on the system's characteristics, like the length and the energy density, is the key feature. If the system is an astrophysical object, like a cluster of galaxies, a black hole, etc, non-negligible corrections arise to several observable quantities.
Distinctive footprints could refer to luminosity distance and strong/weak lensing measurements, among others. The present study focuses on the SNIa luminosity distance observable.

\end{abstract}
\maketitle
%
%
%
%
%

\section{Introduction}

Many high energy theories and phenomenological models include a Cosmological Constant (CC) in their gravity sector, in the most general case, a varying CC, or terms behaving like vacuum density. Based on this CC of geometric or quantum field origin, several studies claim the ability to explain dark energy or inflation by overcoming or ignoring the well-known zero point energy issue or the contribution to curvature from possible cosmic phase transitions. 

The present work refers to models with varying CC but of a specific type. A varying cosmological constant, $\Lambda(x_i,t)$, may be a function of the physical characteristics of the system under consideration, like its spatial size or energy density. This is realized in some existing interesting approaches to quantum gravity. It generates distinctive phenomenological consequences if the system under study is not the whole Universe but an astrophysical object like a cluster of galaxies or a black hole. This work proposes a way of observing these types of $\Lambda$ varying models.

Many high energy and gravitational models could host more or less the idea presented in this letter \cite{Cuzinatto:2022dta,Bondarenko:2019nwv}. In principle, several models that try to explain dark energy are relevant. First, several interesting theoretical works describe and evaluate quantum corrections to the gravity sector due to quantum vacuum energy and field fluctuations, \cite{Bonanno:2020bil},
\cite{Platania:2020lqb}, \cite{Moreno-Pulido:2023ryo}. These types of studies are fundamental, and further work is needed for improvement. In addition, there are studies of models with fluids and associated particles (for example, scalar fields) in a cosmological setup \cite{Motta:2021hvl,Akbarieh:2020hxy}. These models require the use only of well-known quantum field theory, but their validity relies on the existence of new particle fields. Finally, there are the modified gravity models \cite{Bernardo:2022cck},\cite{Mustafa:2023pko},\cite{CANTATA:2021ktz}, which include terms that effectively act similarly to a varying CC. Models of particular interest are screened modified gravities, \cite{Brax:2021wcv, Sakstein:2018fwz, Brax:2012bsa}, like symmetron \cite{Hinterbichler:2010es, Hogas:2023vim, Perivolaropoulos:2022txg, Dong:2013swa, Davis:2011pj}, and chameleon models  \cite{Brax:2004qh, Elder:2023oar, Paliathanasis:2023ttu, Tamosiunas:2022tic, Cai:2021wgv, Karwal:2021vpk, Lombriser:2014dua, Zanzi:2015cch, Burrage:2017qrf}.

\section{Theory}

In this section, we discuss indicative theoretical models that support or differ somewhat from the specific type of scale dependence $\Lambda$ under consideration in this study. In this work, we refer to a $\Lambda$ as a function of length/energy density, and we focus on astrophysical objects and not the whole Universe.

\subsection{Asymptotic Safety}
RG approaches to quantum gravity like Asymptotic Safety (AS) fully support and provide a framework for the required $\Lambda$ behavior discussed in this work, \cite{Eichhorn:2022gku,Knorr:2022ilz,Platania:2022gtt,Bonanno:2020bil,Percacci:2017fkn}.
Quantum corrections (non-perturbative) result in an effective gravity action with a varying cosmological constant, $\Lambda$, which depends on the characteristics of the system under consideration. 

RG methods can help one understand and explore two major issues; UV completeness and how to run from a possibly scale-symmetric regime in the UV to low phenomenology in the IR, \cite{Ahn:2011qt}.
RG functional methods can provide answers to the previous investigations because they go beyond perturbation theory. 
In summary, the functional renormalization group (FRG)
relies on a scale-dependent effective action $\Gamma_k$.
The scale $k$ runs from $\Gamma_{k\to\infty}$ with no quantum fluctuations integrated out, to $\Gamma_{k\to0}$ when all quantum fluctuations have been integrated away.
The FRG flow equation for $\Gamma_{k}$, reads \cite{Wetterich:1992yh, Reuter:1996cp}
\begin{equation}
\label{eq:floweq}
k\partial_k\,\Gamma_k=\frac{1}{2}\mathrm{sTr}\left[\left(k\partial_k\,\Regk\right)\left(\Gamma_k^{(2)}+\Regk\right)^{-1}\right]\,.
\end{equation}
Here, the right-hand side integrates quantum fluctuations with momenta of the order of $k$ contributing most to the change of $\Gamma_k$ at $k$. The second order, $\Gamma_{k}^{(2)}$, a functional derivative with respect to all fields, appears in Eq.(\ref{eq:floweq}), the regulator functional $\Regk$ and the supertrace $\mathrm{sTr}$ that sums over all indices. 

Many gravity actions without and with matter exhibit a non-Gaussian fixed point in the theory space, ensuring UV completeness \cite{Ferrero:2021xqg}. 
The Einstein-Hilbert truncation of the Eﬀective Average Action, used first to demonstrate Asymptotic Safety, see for example \cite{Reuter:1996cp,Reuter:2001ag,Lauscher:2001ya}, is 
\begin{equation}
	\Gamma_k = \Big(16\pi G(k)\Big)^{-1}\bigintsss \di ^4x \sqrt{g}\Big(-R+2\Lambda(k)\Big) + \text{gauge fixing and ghost terms}
	\label{trunc}
	\end{equation}
It results in RG equations with several interesting trajectories and a non-Gaussian fixed point, the so-called Reurter fixed point.
The RG scale dependence $k$ is different from the physical scale dependence. However, the structure of
asymptotically safe models through momentum-dependent correlation functions and form factors lead to the appearance of running couplings that for the gravity sector concern Newton's constant and the cosmological constant, $G(k)$, $\Lambda(k)$. Finally, observables are successfully defined based on these elements of the theory.
A recent paper also proved that the physical graviton propagator is very close to the propagator used for calculations in AS. So the physical running of $G$ and $\Lambda$ is justified \cite{Bonanno:2021squ}.

In the context of AS, the RG flow of  $G(k)$ and $\Lambda(k)$ in the UV regime ($k\rightarrow\infty$) is described by 
\begin{equation}\label{uv}
{G(k)}_{{UV}} = \frac{g_{*}}{k^{2}},  \quad\quad
\Lambda\left( k \right)_{{UV}} = \ \lambda_{*}k^{2},  
\end{equation}
with the dimensionless $(g_{*},\lambda_{*})$  taking finite values (UV fixed point). This indicates an anti-screening of gravity and it can describe the physics near the Big Bang or at the final stages of gravitational collapse.  Eqs.~(\ref{uv}) are valid only in the trans-Planckian regime. A system of partial differential equations describes their evolution. The running behavior of these coupling constants depends on the matter content.
Following the results of the scalar theory, we expect the appearance of an RG trajectory that realizes a tree-level renormalization  \cite{Ohta:2021bkc}.
At lower energies, the dimensionless Newton constant and cosmological constant
 $(g, \lambda$) are running towards a possible IR fixed point respecting the flow
$
g\left( k \right) = g_{*} + h_{1}k^{\theta_{1}}, \quad
\lambda\left( k \right) = \ \lambda_{*} + h_{2}k^{\theta_{2}}, 
$
with $\theta_{1}$, $\theta_{2}$ two positive critical exponents \cite{Bonanno:2005mt}. 
Equivalently for the dimensionful quantities we get
\begin{equation}\label{ir1}
{G(k)}_{{IR}} = \frac{g_{*}}{k^{2}} + h_{1}k^{\theta_{1} - 2}, \quad\quad
\Lambda\left( k \right)_{{IR}} = \ \lambda_{*}k^{2} + h_{2}k^{\theta_{2}+2}.
\end{equation}
Eqs.~(\ref{ir1}) are valid in the infrared regime and they are possibly relevant at astrophysical scales. The parameters that are shown in these equations cannot have any value. For acceptable phenomenology of the late cosmology era, they may obtain the range of values explained in \cite{Zarikas:2017gfv,Anagnostopoulos:2018jdq,Anagnostopoulos:2022pxa}.
Here, $k$ should be understood as the inverse of a characteristic spatial length over which the system's fields are averaged. Thus, $k$ is related to the size of the physical system under consideration. For an astrophysical object, $k$ can be a function of its proper length and thus equivalently of its energy density \cite{Bonanno:2005mt,Bonanno:2017pkg}.  An advanced proper treatment requires the non-perturbative calculation of a finite temperature effective action that contains the gravity field as well as the rest of the fields like those of the Standard model. In our study, we set a certain average value of the CC for all voids to simplify the modeling. We also assume an average energy content and thus a different value of CC for all clusters of galaxies. 

At this point, it is useful to note that the dependence of $G_N$ and $\Lambda$ on the energy density, in the context of AS, of the system under consideration has, of course, implications in dense regions of the Universe, like the early Universe in the "Big Bang" regime, \cite{Kofinas:2016lcz, Bonanno:2009nj, Bonanno:2010mk}, and at the centers of compact objects like stellar interiors, neutron stars, and black holes, \cite{Torres:2014gta, Bonanno:2019ilz, Eichhorn:2022bgu, Kofinas:2015sna}.

In the present study, we have used two values for the Hubble rate. One value is the local expansion rate in low-density systems like voids $H_v$ and a different local expansion rate $H_c$ for the overdense cosmic regions, the filaments/clusters of galaxies.

This study treats the value of Newton's constant $G$ as approximately constant. This also agrees with phenomenologically viable families of RG flow towards the IR limit, where $G$ is a weakly varying quantity.

\subsection{Modified Einstein Equations with $G$ and $\Lambda$ functions of spacetime}

Study, reported in \cite{Bonanno:2020qfu} generalizes previous research conducted in \cite{Kofinas:2015sna} and \cite{Kofinas:2016lcz}. Authors derived modified Einstein equations to accommodate the scenario where the gravitational constant $G$ and the cosmological constant $\Lambda$ vary with respect to spacetime. This subsection provides a concise overview of the obtained results and the underlying assumptions made for this fully consistent and covariant-modified gravity given at the level of equations of motion.
Following the concept of RG approaches to Quantum Gravity like AS, the study introduces a spacetime-dependent cosmological constant $\Lambda(x)$ and Newton's constant $G(x)$ into the 4-dimensional Einstein equations, which then could be 
\begin{equation}
G_{\mu\nu}=-\Lambda(x) g_{\mu\nu}+8\pi G(x) T_{\mu\nu}.
\end{equation}
However, this modification needs more terms to be included to satisfy the Bianchi identities  $G_{\mu\nu}{^{;\mu}}=0$. This can be understood easily since from the Bianchi identities; we get the expression
\begin{equation}
 8\pi(GT_{\mu\nu}){^{;\mu}}=\Lambda_{;\nu}.   
\end{equation}
However, this expression presents a problem since, in the absence of matter ($T_{\mu\nu}=0$), it yields a constant value for $\Lambda$. To consistently support the variation of $G$ and $\Lambda$ in spacetime, it becomes necessary to include additional covariant derivatives of $G$, $\Lambda$, and $T_{\mu\nu}$ in the Einstein equations. It is important to note that in this study \cite{Bonanno:2020qfu}, $G$ and $\Lambda$ are not considered independent fields, unlike in models such as Brans-Dicke theory, and thus, they lack their own equations of motion. This aligns with RG approaches to Quantum Gravity, such as Asymptotic Safety, where $G$ and $\Lambda$ are treated as running coupling constants within an effective field theory framework. 
Another assumption made in \cite{Bonanno:2020qfu} is that both $G$ and $\Lambda$ do not change signs during their evolution, which is evident for $G$ but restrictive for $\Lambda$. The rationale behind this assumption is supported by various RG studies, which indicate the preference for $\Lambda$ to always be positive or negative.

A constraint is imposed to construct the modified equations further, requiring that the additional kinetic terms vanish when $G$ and $\Lambda$ are constants to recover the Einstein equations. This implies that these extra terms must include at least one covariant derivative of $G$ or $\Lambda$. Additionally, it was assumed that the coefficients of the kinetic terms should be determined from the Bianchi identities. This rules out including terms containing $T$ or $T_{\mu\nu}$. Thoroughly examining various combinations of tensorial terms it was showed that the extra kinetic terms should have an even number of covariant derivatives for $G$ and $\Lambda$. For simplicity, up to second-order derivatives have been included in the modified equations,
\begin{equation}
G_{\mu\nu}=-\Lambda g_{\mu\nu}+\vartheta_{\mu\nu}+\widetilde{\vartheta}_{\mu\nu}
+\mathcal{F}(\psi_{;\mu}\chi_{;\nu}+\psi_{;\nu}\chi_{;\mu})
+\mathcal{H}g_{\mu\nu}\psi^{;\rho}\chi_{;\rho}+8\pi G T_{\mu\nu}\,,
\label{njgc}
\end{equation}
with
\begin{equation}
\vartheta_{\mu\nu}=\mathcal{A}\,\psi_{;\mu}\psi_{;\nu}+\mathcal{B}\,g_{\mu\nu}\psi^{;\rho}
\psi_{;\rho}+\mathcal{C}\,\psi_{;\mu;\nu}+\mathcal{E}\,g_{\mu\nu}\Box\psi\,,
\label{fsdmn}
\end{equation}
\begin{equation}
\widetilde{\vartheta}_{\mu\nu}=\widetilde{\mathcal{A}}\,\chi_{;\mu}\chi_{;\nu}+
\widetilde{\mathcal{B}}\,g_{\mu\nu}\chi^{;\rho}\chi_{;\rho}
+\widetilde{\mathcal{C}}\,\chi_{;\mu;\nu}+\widetilde{\mathcal{E}}\,g_{\mu\nu}\Box\chi
\label{fsdjn}
\end{equation}
where $\Lambda=\bar{\Lambda}\,e^{\psi}$ and $G=\bar{G}\,e^{\chi}$ and $\bar{\Lambda}$ and $\bar{G}$ being arbitrary constants,
and all coefficients $\mathcal{A}, \mathcal{B}, \mathcal{C}, \mathcal{E}, \widetilde{\mathcal{A}},
\widetilde{\mathcal{B}}, \widetilde{\mathcal{C}}, \widetilde{\mathcal{E}}, \mathcal{F},
\mathcal{H}$ are functions of both $\psi, \chi$.

Finally, it was proved that the modified Einstein equations have uniquely defined second-order extra kinetic terms :
\begin{equation}
G_{\mu\nu}=-\bar{\Lambda}\,e^{\psi} g_{\mu\nu}
-\frac{1}{2}\psi_{;\mu}\psi_{;\nu}-\frac{1}{4}g_{\mu\nu}\psi^{;\rho}
\psi_{;\rho}+\psi_{;\mu;\nu}-g_{\mu\nu}\Box\psi+8\pi G T_{\mu\nu}
\label{eq:field_equations}
\end{equation}
while the conservation equation is
\begin{equation}
\big(GT_{\mu\nu}\big)^{;\mu}\!+\!G\Big(T_{\mu\nu}\!-\!\frac{1}{2}Tg_{\mu\nu}\Big)
\psi^{;\mu}=0\,.
\label{eq:conservation equation}
\end{equation}

This set of equations is an example of another theoretical framework that can host the concept of our study. Assuming now results form Asymptotic Safety, \cite{Bonanno:2005mt},\cite{Bonanno:2017pkg}, or similar RG approaches, we can set for the astrophysical energy scales
\begin{equation}
\Lambda= c1\,\rho^{1/2},\,\,\,\,\,  G\approx G_N.      
\end{equation}
where $c1$ is a constant and $\rho$ is the energy density of the system under consideration.
To proceed further and solve the new Einstein and conservation equations, Eqs. (\ref{eq:field_equations}) and (\ref{eq:conservation equation}) should be used working with a spatially homogeneous and isotropic metric \cite{Bonanno:2020qfu}, or an inhomogeneous metric or spherical symmetry to derive spherical solutions.

\subsection{Running Vacuum Models}
There is a particularly interesting class of models under the name Running Vacuum Models (RVM), \cite{SolaPeracaula:2022hpd, Moreno-Pulido:2022phq, GomezValent:2017kzh}. There is an RG running of $\Lambda$ in these models but a perturbative one that differs from the non-perturbative approach in AS.
Their theoretical foundation is based on applying quantum field theory in curved spacetime. Working with a scalar field, it is possible to calculate its zero point energy (which contributes to the cosmological constant) and avoid infinities taking a re-normalization difference prescription scheme for the zero point energies for two different energy scales and at two different cosmic epochs. The action used in RVM is
\begin{equation}\label{eq:Sphi}
  S[\phi]=-\int d^4x \sqrt{-g}\left(\frac{1}{2}g^{\mu \nu}\partial_{\mu} \phi \partial_{\nu} \phi+\frac{1}{2}(m^2+\xi R)\phi^2 \right)\,.
\end{equation}
where $\xi$ is assumed arbitrary. CC in these models is a function of energy scale and the expansion of the universe. 
Working in a flat homogeneous and an isotropic background spacetime, it has been proven \cite{SolaPeracaula:2022hpd} that the CC is a function of two energy scales $M$, $M_0$ and two different cosmic epochs characterized by their expansion rate $H$. $\Lambda=f(M,M_0,H, H_0)$ is given by the following expression
\begin{equation}\label{rho}
\begin{split}
\Lambda(M,H)-\Lambda(M_0,H_0)&
=\frac{3\left(\xi-\frac{1}{6}\right)}{2\pi}\,G\left[H^2\left(M^2-m^2+m^2\ln\frac{m^2}{M^2}\right)\right.\\
&\left.-H_0^2\left(M_0^2-m^2+m^2\ln\frac{m^2}{M_0^2}\right)\right]+\textit{O} \left(H^4\right)\,.
\end{split}
\end{equation}

Following our proposed way to seek observational evidence, we must consider local values of $\Lambda$ and not work with a global cosmic value.
Thus, we choose as $M_0$ and $H_0$ the energy scale and the expansion rate today, and consequently, we assign a value $M_v$ in the case of voids and a different value $M_c$ for the case of clusters since they are systems of different energy content. So, in these models, the calculations of luminosity distance are affected and are different from the conventional ones. It is a similar case to the one we study here but differs somewhat and is more technically complicated. The expression for the luminosity distance we present in section IV does not cover this case. It is beyond the scope of the present study to analyze this class of models. The previous two models in this section are more simple cases because the $\Lambda$ is a function only of the energy scale.
The next section focuses on the distinctive footprints concerning luminosity distance measurements.

\section{Clusters of galaxies/filaments and Voids}

These RG flow methods to explore quantum gravity are still under development and require further elaboration.  
However, the present study does not rely on details of the RG flow. The proposed phenomenological signature assumes only a varying cosmological constant whose values are different if it refers to different astrophysical objects with different sizes and energy content. 

Regarding the observable implications, the setup concerns the clusters of galaxies in filaments and voids. For the reasons we have explained, we propose that the clusters of galaxies in filaments are associated with a cosmological constant of a different value than the value attributed in cosmic voids.




At this point, we have to state some interesting possibilities to proceed:

\textbf{Case 1}. Perhaps this is the most expected case from the AS phenomenology point of view.
Cosmic voids (low-density systems) are described by a gravity action that includes a negative or zero or negligible positive CC while regions of space with clusters of galaxies are associated with positive values of CC of the order of the value used for the $\Lambda$CDM model. The exact value of a CC, on the astrophysical scale, is not known; but working with a positive value like in \cite{Zarikas:2017gfv, Anagnostopoulos:2022pxa}, it has been proved that it is possible to explain the recent cosmic acceleration without any fine-tuning, something that is unique to the best of our knowledge. Studies \cite{Zarikas:2017gfv, Anagnostopoulos:2022pxa}, have as a working assumption a positive CC with a value related to the astrophysical length of the cluster of galaxies.

\textbf{Case 2}. 
Due to the absence of widely accepted concrete IR quantum gravity corrections, it is useful to mention also another possibility that can be phenomenologically interesting.
A reversed setup could be that now in low densities systems like voids, there is a positive  CC of the order of the value that is compatible with the $\Lambda$CDM while in higher densities systems like regions of space with clusters of galaxies, we have either a negligible positive CC or a negative CC which may have a value that could contribute to the missing dark matter.
Or, it could be that there is a negligible positive or negligible negative CC in low-density systems like voids, while in higher densities systems like regions of space with clusters of galaxies, we have a negative cosmological constant that could contribute to the missing dark matter.

Now, the question that arises is how to determine approaches to dig out observational support for this variation of 
$\Lambda$.
One simple approach is to utilize signals from two samples, one related to signals passing through voids and one for signals passing via clusters of galaxies/filaments. Thus, measuring a statistically significant difference in the mean value of $\Lambda$, via a statistical methodology and choosing appropriate observables would be a footprint indicating new physics. This approach is a neat method, and it needs very detailed information about the sources, perhaps available in future observational data. Another approach would be to explore $\Lambda$ variations considering signals coming from different cosmic regions, cosmic parts with under-densities and over-densities.

In this study, we will elaborate on the phenomenological consequences of these two cases for the case of Supernovae distance measurements. 
Thus, we will assume different average values of the $\Lambda$ between regions of spacetime without matter (or very low energy density) and regions of the Universe with matter like filaments with clusters of galaxies, and we will evaluate how the luminosity distance is affected.



\section{Observational signatures}
The observable to be studied is the luminosity distance. For a globally homogeneous universe (assuming voids and walls with galaxy clusters distributed homogeneously), the metric is 
\begin{equation}
ds^2=-c\,dt^2+a^2(t)\left( \frac{1} {\sqrt{1-k\,r^2}} \,dr^2+ r^2\,d\Omega^2 \right),
\end{equation}
with 
\begin{equation}
  d\Omega^2=d\theta^2+sin^2\theta\,d\phi^2.
\end{equation}
A light ray coming towards us travels the proper distance
\begin{equation}
  d=c\,dt=a(t) \frac{dr}{\sqrt{1-k\,r^2}},  
\end{equation}
and from now on, we set $c=1$. It is useful to calculate the comoving distance $S(r)$
\begin{equation}
S(r) = \int_0^r \frac{dx}{\sqrt{1-k\,x^2}}  ,
\end{equation}
with the well-known result that  
\begin{equation}
S(r) = \left\{  \begin{array}{ll}
      \arcsin (r) & k=1 \\
      r & k=0\\
      \arcsinh(r) & k=-1   \nonumber \\ \end{array}  \right. .
\end{equation}
The Hubble rate in general is $ H=\dot{a}/a=\frac{-1}{(1+z)}\frac{dz}{dt} $ which is written as
\begin{equation}\label{H}
H^2 = H_0^2 \left[ \Omega_M\,(1+z)^3 +\Omega_\gamma\,(1+z)^4 + \Omega_\Lambda + \Omega_k\,(1+z)^2 \right] .       
\end{equation}
The several $\Omega$ that appear in Eq.~(\ref{H}) represent the matter, radiation, cosmological constant, and topological curvature part.

By definition, the luminosity distance is $D_L=\sqrt{L/4\pi F}$. The rate of photons is reduced by a factor $a/a_0=1/(1+z)$, and there is also a further energy reduction by the same factor due to redshift. Note that we set $a(t=0)=1$ as usual. Finally $L_{obs}=a(t_{emit})^2\,L_{emit}$ and since
\begin{equation}
F=a^2\frac{L}{4\pi\,S(r)^2},
\end{equation}
the luminosity distance is
\begin{equation}
D_L = (1+z)S(r).
\end{equation}
Another useful equivalent expression that gives the luminosity distance is 
\begin{equation}\label{DL}
D_L = -(1+z)\int_{t_0}^{t_{emit}}\frac{dt}{a(t)}=-(1+z)\int_{a}^{1} \frac{da}{a^2\,H(a)} =(1+z)\int_{0}^{z} \frac{dz'}{H(z')} ,
\end{equation}
with $z$ the redshift of the source.

\subsection{Large-scale structure map approach}

A simple way to model the astrophysical setup is to work with a different scale factor $a(t)$ and consequently Hubble rate in the voids compared to the ones used for the regions of space with matter due to the local spatial variations of CC. Of course, this is a simple model, and a fully inhomogeneous treatment is needed. However, although this model is an approximation, it clearly presents the idea to the reader, which is the main aim of this study.
In the relevant integral, Eq.~\ref{DL} we have to integrate from the redshift of the distant object till today. The integral will then split into several parts depending on the path. 

Let us suppose we have a path that initially begins from a distant cluster region after passing from a void and after coming into our cluster region; then, the integral will be split into three parts where the scale factor and the Hubble rate will be different.
Then, the total luminosity distance is,

\begin{eqnarray}\label{DLV}
D_L &=&\frac{a_{0}}{a_{vf}} \int_{0}^{Zvf} \frac{dz'}{H_{c}}+ \frac{a_{vf}}{a_{vi}}\int_{Zvf}^{Zvi} \frac{dz'}{H_{v}}+\frac{a_{vi}}{a_{emit}}\int_{Zvi}^{Z} \frac{dz'}{H_{c}}, \nonumber \\
&=& ({1+z_{vf}}) \int_{0}^{Zvf} \frac{dz'}{H_{c}}+ \frac{1+z_{vi}}{1+z_{vf}}\int_{Zvf}^{Zvi} \frac{dz'}{H_{v}}+\frac{1+z}{1+z_{vi}}\int_{Zvi}^{Z} \frac{dz'}{H_{c}},
\end{eqnarray}
where $z_{vi}$, $z_{vf}$, $z$ the redshifts at the first entrance of the signal into the void, the redshift of the signal at the exit from the void, and the redshift of the source respectively.
In addition for a region with clusters of galaxies, we have
\begin{equation}\label{Hc}
H_{c}^2 = H_0^2 \left[ \Omega_M\,(1+z)^3  + 
\Omega_{\Lambda c} + \Omega_k\,(1+z)^2 \right].       
\end{equation}
This formula is, of course, valid for a Universe that consists only of homogeneously distributed voids and clusters.  
In analogy, for voids one has,
\begin{equation}\label{Hv}
H_{v}^2 = H_0^2 \left[ \Omega_\gamma \,(1+z)^4 +  \Omega_{\Lambda v}+\Omega_k\,(1+z)^2 \right].        
\end{equation}

Now, in the case the signal passes from $N$ voids and $M$ clusters/filaments, the luminosity distance is
\begin{eqnarray}\label{DLNΜ}
D_L = (1+z_{vf\,N}) \int_{0}^{Zvf\,N} \frac{dz'}{H_{c}}+ \sum_{p}\,\frac{1+z_{vi\,p}}{1+z_{vf\,p}}\int_{Z_vf\,p}^{Zvi\,p} \frac{dz'}{H_{v}}+ \nonumber \\
+ \sum_{q}\,\frac{1+z_{ci\,q}}{1+z_{cf\,q}}\int_{Z_cf\,q}^{Zci\,q} \frac{dz'}{H_{c}}+\frac{1+z}{1+z_{vi\,1}}\int_{Zvi\,1}^{Z} \frac{dz'}{H_{c}},
\end{eqnarray}
with $p=N,N-1,...1$ and $q=M-1,M-2,...2$ and where $z_{vi\,p}$, ($z_{ci\,q}$), $z_{vf\,p}$, ($z_{cf\,q}$), the redshifts at the first entrance of the signal into the p-th void (q-th cluster) and the redshift of the signal at the exit from this void (cluster) respectively.

Regarding statistics for these suggested cases that concern the measurement of distance, we can test several different null hypotheses (working with the luminosity distance as an observable). 

1) Statistical comparison of the mean value of the observable, $D_L$, evaluated from observations with the theoretical value of the same observable if we assign a favored from theory value for $\Lambda$. Since we can estimate $D_L$ from Eq.~\ref{DLNΜ} or from another similar approach and at the same time, we estimate distances from the magnitudes of SN Type Ia (in the conventional way), we can compare these two $D_L$.

2) To compare the mean values of the $\Lambda$ from two samples statistically. One sample contains signals that had passed through certain numbers of voids and clusters, and one different sample with signals that have passed through the same numbers of clusters and voids.  Performing calculations for all signals that belong to the same sample, we will calculate several values for the $\Lambda$ and compare their mean/median with the mean/median of the values calculated from the other sample. Assuming that all voids and clusters are roughly similar sizes, we have to find no statistical difference in the mean/median value of $\Lambda$ from the two samples. This test is meaningful as a pair with the third test.

3) Another test could be the following: construct one sample with signals whose line of sight contains much more voids than the second sample of signals. Then we have to find statistical differences in the mean value of the $\Lambda$ between the two samples. This would be an encouraging indication for quantum corrections to $\Lambda$.

4) Another obvious test could be to check which models fit best the SN1a data, the $\Lambda$CDM or the model with $\Lambda$ variation proposed in the present work (check all two cases separately). For this type of test, signals from SN events included in the sample should pass from a known number of voids and fillaments with clusters.  


The quantity $H_0$ is supposed to be known, so it is fixed to one value for all these tests.

\subsection{Under-densities and over-densities approach}
This approach is more feasible and possible within future levels of observations. 
If it is not possible to know the details about the distances between voids and clusters, then we can work as follows. We can define two samples, one in which voids dominate and one in which clusters/filaments dominate.
So we select a large sample of signals with line-of-sight through regions with over-densities and another sample from regions with under-densities. However, the larger the
sample, the smaller the difference between the densities. There must be a trade-off.

Suppose we get signals from a sample (sample 1) of astrophysical sources that arrive on earth passing from an under-dense cosmic region. Let us assume that this region has an average matter density given by  
\begin{equation}
\rho_1=\epsilon_1\,\rho_{cr},   
\end{equation}
where $\rho_{cr}$ is the critical mass density of the universe and $0<\epsilon_1<1$. Then, a second sample should be determined with signals passing regions with over-density i.e., with an average matter density $\rho_1=\epsilon_2\,\rho_{cr} $ with  $\epsilon_2>1$. Since these two coefficients $\epsilon_i$, with $i=1 \,\text{or}\, 2$, can be known, consequently, we can search for statistical differences in the values of $\Lambda_i$.

The relevant integral \ref{DLNΜ} does not split into parts. The luminosity distance is given by
\begin{equation}\label{DLrho}
D_{L i}  = (1+z)\int_{0}^{z} \frac{dz'}{H_{i}(z')} ,
\end{equation}
with 
\begin{equation}
H_{i}^2 = H_0^2 \left[ \Omega_M\,(1+z)^3  + \frac{\Lambda_i}{3H_0^2} + \Omega_k\,(1+z)^2 \right],        
\end{equation}
with cosmological constant given according to the AS theory, \cite{Bonanno:2005mt},\cite{Bonanno:2017pkg},  from the expression
\begin{equation}
\Lambda_i= \xi\,\rho_i^{1/2}.       
\end{equation}
We can also further test if the following formula is true for the samples' two evaluated values of $\Lambda$.
\begin{equation}
\frac{\Lambda_1}{\Lambda_2}=\frac{\rho_1^{1/2}}{\rho_2^{1/2}}=\frac{\epsilon_1^{1/2}}{\epsilon_2^{1/2}}.
\end{equation}
This last test would be an important indication of AS or similar RG approaches to quantum gravity.

\subsection{Constraints}

We used two indicative values for the Hubble rate in the present study. One value is the local expansion rate in low-density systems like voids $H_v$ and a different local expansion rate $H_c$ used for the over-density cosmic regions, the filaments/clusters of galaxies.
There are known constraints in the amount of spatial variability for the $\Lambda$ that causes anisotropies and inhomogeneities in some cosmic observables.
For example, constraints come from the Cosmic Microwave Background \cite{Appleby:2009za}. However, AS models that we mainly discuss in our work overpass this problem since the spatial variability under discussion appears in late cosmology after the large-scale structure.
Relevant constraints arise from late cosmology data. In \cite{Bahr-Kalus:2012yjc}, data from supernovae SNIa analyze hemispherical asymmetries and find a maximum allowed variation on the Hubble rate $\frac{\Delta H}{H}<0.03$. Thus, in our modeling, we must respect that
\begin{equation}
 \frac{\lvert H_c(z=0)-H_v(z=0))\rvert}{H}<0.03 .  
\end{equation}
This is a constraint that can be achievable in the context of AS for a set of free parameters. Note also that this is a strict bound, and it will probably be relaxed if someone follows our approach for sampling. The reason is that in all papers concerning constraints that appear in the literature, $H$, is computed from a luminosity distance with the conventional way.

Note that in the present work, we don't claim that the different values of $\Lambda$ in voids and filaments are such that they necessarily explain the recent cosmic acceleration partially or fully or explain other cosmological tensions, \cite{Abdalla:2022yfr}. This paper suggests a treatment of observations to verify this quantum gravity origin scale dependence of $\Lambda$.

The amount of the position dependence or large scatter in standardized light curves of supernovae can vary depending on several factors, including the specific sample of supernovae being studied, the quality of the data, and the analysis techniques employed. Scatter can vary from case to case. 

\textit{Intrinsic Scatter}: Supernovae of different types or subtypes can exhibit significant variations in their light curves, even when standardized. This intrinsic scatter arises from differences in explosion mechanisms, progenitor properties, and other intrinsic factors. For example, Type Ia supernovae, commonly used as standard candles, still exhibit variations in their light curves, albeit with a narrower scatter than other types.
\textit{Extrinsic Scatter}: The presence of extrinsic factors, such as dust extinction, line-of-sight effects, or host galaxy properties, can introduce additional scatter in the standardized light curves. The impact of these factors can vary depending on the specific observational conditions and the properties of the host galaxies.

In \cite{Vinko:2018crs}, authors report a variation in the estimations of absolute distances due to different methodologies. They report that comparing the distance estimators with the various codes reveals important
constraints on the systematics present either in the basic assumptions of the methods or in their implementation and calibrations. For the best-fit parameters provided by the different codes, MLCS2k2, SNooPy2 and SALT2.4, the distance moduli of moderately reddened SNIa agree within $\leq 0.2$ mag. This bound refers to the distance moduli which have dimensionality of magnitude and can be transformed to luminosity distance $D_L$, which we are using in the present study, with the known formula $ m - M = 5 \log(D_L / 10 pc) $. One outcome of the work \cite{Vinko:2018crs}, is that this uncertainly in distances gets larger for more distant supernovae signals. This constraint comes from SN light curve fittings that work with a homogeneous cosmology. We expect the dispersion will be larger, working under our hypothesis.
The present work proposes a method to discover a $\Lambda$ dependence on the energy density of the system measuring distances from SNIa. We don’t know from theory exactly how much the difference of the value of $\Lambda$ is between clusters/filaments and voids, but almost certainly, RG approaches to quantum gravity imply a position dependence on the estimation of distances using SNIa due to the special type of inhomogeneities in $\Lambda$, i.e. different values of $\Lambda$ in overdense and underdense regions.  
The inhomogeneities in $\Lambda$ that we describe in our work are unlike CMBs'. They cannot be modeled, for example, with a scalar field with spatial variations like those in paper \cite{Dutta:2006pn}.
They are strictly related to the recent large-scale structure. Since many signals pass through underdense and overdense regions, they experience an average $\Lambda$. So sampling is very important if we want to discover the proposed effect.
None, to the best of our knowledge, so far distinguished a sample that contains SNIa signals that pass only from overdense regions and another sample that passes from underdense regions; only then will it be possible to discover differences. Sampling is crucial in the context of supernovae; for example, in the paper \cite{NearbySupernovaFactory:2018qkd}, a special sampling revealed Local stellar mass bias. 

Another relative issue is that Type Ia supernovae have light curves that have widths and magnitudes that can be used for testing cosmologies. However, standard analysis calibrates the light curve against a rest-frame average (such as SALT2). Consequently, type Ia supernovae calibrated with these methods cannot be used to investigate in-homogeneous models of cosmologies. 

There will be several difficulties in observing statistical differences using SNIa distance measurements after the sampling. Various factors have to be considered seriously.

The calibration of the absolute brightness of SNIa (luminosity) is crucial for distance measurements. Uncertainties in the calibration process lead to errors in determining their true luminosity. Furthermore, the light from SNIa can be absorbed and scattered by dust in the interstellar medium of their host galaxies or along the line of sight. Correcting for this dust extinction introduces quite important uncertainties, especially if the dust properties are not well understood. These uncertainties must be smaller than the difference in the mean/median value we propose to be detected.

The properties of the host galaxies where SNIa occur can affect their observed brightness. Differences in the star formation history, metallicity, and other galaxy properties can also introduce systematic errors that have to be minimized or avoided. For example, in our case, we could use supernovae that share very similar properties in the proposed two samples. In our case, this does not comprise a selection bias i.e., selection of SNIa candidates for observation of certain types of supernovae or specific host galaxy properties. This is not relevant bias, in our proposal, since we want to find a statistical difference between the two samples. Using the two sample signals from the same observational techniques and software codes would be beneficial too.

Finally, cross-validation with other cosmological probes, such as weak or strong gravitational lensing, could also be used to uncover a possible $\Lambda$ depending on energy density, but this is outside the scope of the present work.

\section{Conclusions}

In this study, we propose a specific observational signature to test $\Lambda$ varying gravity models inspired by quantum gravity.
If $\Lambda$ varies due to quantum corrections and becomes a function of a characteristic length or equivlently energy density of the matter configuration under consideration, there are astrophysical and cosmological consequences. The luminosity distance is taken as an observable, and a method was proposed to explore the possibility that $\Lambda$, takes different values in cosmic voids compared to the values in clusters/filaments.

As a future work, a more thorough analysis is needed using inhomogeneous cosmological metrics. 
Several interesting works evaluate the luminosity distance in inhomogeneous cosmological models \cite{Helbig:2019jcm}.
However, in all these works that include a CC, this CC takes the same value in the different parts of the Universe.
Thus, new solutions should be found and  consequently improved expressions for the luminosity distance.  Furthermore, simulations would also be useful to be performed for each of the two different cases presented in this study.

\section{acknowledgments}
V.Z. acknowledges enlightening discussions with Patrick Petitjean. Funding comes in part from the FY2021-SGP-1-STMM Faculty Development Competitive Research Grant No. 021220FD3951 at Nazarbayev University. 




\newpage
\bibliography{ref.bib}

\end{document}